\documentclass[prl,twocolumn]{revtex4}
\usepackage{graphicx}
\renewcommand{\narrowtext}{\begin{multicols}{2} \global\columnwidth20.5pc}
\newcommand{\pref}[1]{(\ref{#1})}
\renewcommand{\v}[1]{{\bf #1}}

\newcommand{\gr}{{\nabla}}
\def\be{\begin{eqnarray}}
\def\ee{\end{eqnarray}}
\newcommand{\nn}{\nonumber\\}
\newcommand{\Eq}[1]{Eq.~(\ref{#1})}

\newcommand{\ra}{\rightarrow}

\begin{document}
\draft

\title{Mott Insulators Without Symmetry Breaking}
\author{Dung-Hai Lee$^{a,b}$, and Jon Magne Leinaas$^{a,c}$}
\affiliation{${(a)}$Department of Physics,University of California
at Berkeley, Berkeley, CA 94720, USA}\affiliation{${(b)}$ Material
Science Division, Lawrence Berkeley National
Laboratory}\affiliation{${(c)}$Department of Physics,University of
Oslo, P.O. Box 1048 Blindern, 0316 Oslo, Norway}

\date{\today}
\begin{abstract}
We present theoretical models, in one and two space dimensions,
that exhibit Mott insulating ground states at fractional
occupations without any symmetry breaking. The Hamiltonians of
these models are non-local in configuration space, but local in
phase space.
\end{abstract}
\maketitle
\parindent 10pt In recent years the study of doped Mott insulators \cite{mott}
has become one of the main themes of condensed matter physics
\cite{revmott}. In a Mott insulator the motion of each particle is
hindered by the presence of other particles due to their strong
repulsive interaction, and in this respect they are different from
band insulators where the Pauli exclusion principle forbids the
motion of electrons. An interesting feature of Mott insulators is
that doping, {\it i.e.}, changing of particle density, can
dramatically affect their properties. For example, doping of
``cuprates'' and ``mangnites'' has lead to the high transition
temperature superconductors and the colossal magneto-resistive
materials.

If a crystalline solid has a fractional occupation number
\cite{note10} and is Mott insulating, there seems always to be a
spontaneous symmetry breaking of the translation symmetry
\cite{thm}. After the symmetry breaking the unit cell is
effectively enlarged so that the occupation number becomes an
integer. This kind of symmetry breaking blurs the distinction
between Mott insulators and band insulators\cite{band}.

Nearly thirty years ago, Anderson and Fazekas \cite{andf}
postulated that due to the frustrated antiferromagnetic
interation, the Mott state on triangular lattice does not break
the spin rotation symmetry (hence is ``featureless''). This novel
Mott state has since been dubbed the name ``spin liquid''. Soon
after the discovery of high temperature superconductivity,
Anderson proposed that the parent (Mott insulating) compounds of
these superconductors are spin liquids \cite{rvb}. Although
subsequent developments have shown this is not true, nonetheless
this proposal has stimulated wide interest in spin
liquids\cite{spinlq}.

 One of the reasons that we are interested in
``featureless'' Mott insulators is because it is widely believed
that the elementary excitations in such  systems may carry
fractional quantum numbers (e.g., fractional charge, fractional
spin, etc). As a result, the conducting state(s) produced by
doping them can be vrey different from the those
produced by doping ordinary symmetry-breaking Mott insulators. 
In this paper we present two examples that exhibit featureless
Mott insulating ground states. In both cases the elementary
excitations carry fractional quantum numbers. From these examples
we learn a valuable lesson in constructing models of this kind.
\\


\parindent 10pt Without further delay let us define these models.
First we focus on a one-dimensional (1D) model. Consider a finite
lattice with $M=3N-2$ sites where $N$ is an odd integer. We label
each site by an integer $n$ with the central site $n=0$. On this
lattice we put $N$ spinless fermions. The occupation number is
given by $\nu=N/(3N-2)$ which approaches $1/3$ in the $N\ra\infty$
limit. It turns out that this choice of the number of sites
ensures that the ground state is non-degenerate for all $N$.  The
Hamiltonian that describes the interaction between these N
fermions is given by \be &&H_{1D}= g\kappa^3 \sum_p b^+_p
b_p.\label{model1}\ee In the above $g$ is an energy parameter, and
is positive. The parameter $\kappa$, whose meaning shall be
discussed latter, is dimensionless ($0\le\kappa<1$). Since
$H_{1D}$ is a sum of positive-definite operators, its eigenvalues
are clearly non-negative.

In $H_{1D}$ the index $p$ runs through both integers and half
integers, and \be b_p={\sum_{q}}\Big(q e^{-\kappa^2
q^2}\Big)c_{p-q}c_{p+q},\label{bn}\ee where $q$ takes
integer/half-integer value depending on whether $p$ is
integer/half-integer. In \Eq{bn} $c_{p\pm q}$ annihilates a
fermion at lattice site $p\pm q$. $H_{1D}$ describes the hopping
of pairs of fermions from sites $p-q$ and $p+q$ to $p-k$ and $p+k$
(Fig.1). In such a hopping process the center-of-mass position
($p$) of the fermion pair remains unchanged. From \Eq{bn} it is
apparent that $\kappa^{-1}$
controls the hopping range.
\begin{figure}
\includegraphics[width=7.5cm,height=1cm]{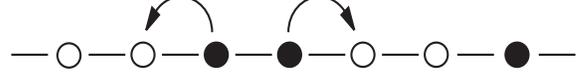}
\caption{The effect of $H_{1D}$ is to hop a pair of particles
while conserving their center of mass position. The hopping range
is set by $\kappa^{-1}$. In this figure the nearest neighbor pair
hopping is illustrated. }
\end{figure}


For occupation $\nu=N/(3N-2)$ and {\it any} $\kappa$, the model
defined by \Eq{model1}
possesses a ground state 
given by \be &&|\Psi_{1D}> =\sum_{\{n_k\}}
\chi(n_1,...,n_N)~c^+_{n_1}...c^+_{n_N}|0>,\label{wf1}\ee where
\be &\chi&(\{n_j\})\propto \prod_j~e^{-{\kappa^2 n_j^2/2}
}\int_{-\infty}^{\infty}dx_j\int_0^{2\pi}dy_je^{-{x_j^2/\kappa^2}}\nn&\times&
e^{n_j(x_j-iy_j)}\prod_{j>k}
\sinh^3\Big[{(x_j-x_k)+i(y_j-y_k)\over 2}\Big].\label{wfn1}\ee
This state is annihilated by all the $b_p$'s, {\it i.e.}, \be b_p
|\Psi_{1D}>=0.\label{annh}\ee As a the result
$H_{1D}|\Psi_{1D}>=0$ implying 
$|\Psi_{1D}>$ is a ground state of $H_{1D}$. \Eq{annh} can be
shown to be true by rewriting it in terms of the wavefunction
$\chi(\{n_j\})$ and using the following identitiy\cite{noteee} \be
&&\int_{-\infty}^{\infty} dx_1dx_2\int_0^{2\pi}dy_1dy_2~
\Phi^*_p(z_1,z_2)\prod_{j>k}\sinh^3\Big({z_j-z_k\over
2}\Big)\nn&&\times e^{-{1\over 2\kappa^2}\sum_k x_k^2}=0,~~{\rm
for~all} ~p.\label{prf}\ee The $\Phi_p$ in \Eq{prf} is given by
\be \Phi_p(z_1,z_2)&&=\Big[ e^{-\kappa^2p^2}e^{p(z_1+z_2)-{1\over
4\kappa^2}(x_1+x_2)^2}\Big]\nn&&\times\Big[\sum_q q
e^{-2\kappa^2q^2}e^{q(z_1-z_2)-{1\over
4\kappa^2}(x_1-x_2)^2}\Big].\ee

In latter part of the paper we shall show that for sufficiently
small positive values of $\kappa$,  $|\Psi_{1D}>$ is the unique
ground state of $H_{1D}$, it separated by an energy gap from the
lowest excited states, and it does not break translational
symmetry. Meanwhile, to get a feeling for $|\Psi_{1D}>$, let us
look at a representation of it for 3 particles, as shown in Fig.2.
The result is a coherent superposition of many different
configurations. The relative weight and phase of these
configurations ensures that they are annihilated by $b_p$. Any
breaking of the weight-phase relation causes excited states.
\begin{figure}
\includegraphics[width=6cm,height=4.5cm]{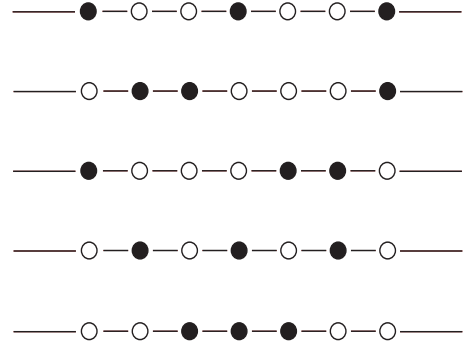}
\caption{$|\psi_{1D}>$ for three particles. It is represented by
components with different occupation of sites. The relative weight
and  phase between these configuration are $e^{9\kappa^2}$ (top),
$3e^{7\kappa^2}$,$3e^{7\kappa^2}$,$6 e^{4\kappa^2}$, $-15
e^{\kappa^2}$ (bottom) respectively. }
\end{figure}

Next we present a two dimensional (2D) model with similar
properties. The lattice consists of $M'$ chains, with each chain
made up of $M=3N-2$ sites. In each chain we place $N$ spinless
fermions. As the result, the 2D occupation number is also given by
$\nu=N/(3N-2)$. In the limit $N,M'\rightarrow\infty$ the
occupation number is also $1/3$. The Hamiltonian of the 2D model
is given by \be H_{2D}&=&\sum_{n\,p}\Big[ g_1\kappa^3 \,
b^+_{(n,p)}b_{(n,p)}+g_2\kappa^3 (~ b^+_{(n,p)}b_{(n+1,p)} \nn
&&+b^+_{(n+1,p)}b_{(n,p)}~)\Big].\label{model2}\ee In the above
the chain index $n$ runs through integers and $p$ can be both
integer and half-integers. In $H_{2D}$ \be
b_{n,p}={\sum_{q}}\Big(q e^{-\kappa^2
q^2}\Big)c_{(n,p-q)}c_{(n,p+q)},\ee  where  $q$ runs through
integer/half-integer depending on whether $p$ is
integer/half-integer. The energy parameters $g_1$ and $g_2$ are
both positive. The first term of $H_{2D}$ hops a pair of particles
within each chain $n$, and the second term hops a pair of
particles from chain $n$ to chain $n+1$ and vice versa. In both
hopping processes the center of mass coordinate along the chain
direction remains unchanged.

For any values of $g_1$ and $g_2$, \be |\Psi_{2D}>=\prod_n
|\Psi_{1D,n}>,\label{st2}\ee where $|\Psi_{1D,n}>$ is the previous
$|\Psi_{1D}>$ defined for the $n$th chain, is a zero-energy {\it
eigen state} of $H_{2D}$. This is because $b_p$ annihilates
$|\Psi_{1D}>$. Let us imagine obtaining $H_{2D}$ by turning on
$g_2$ from zero while holding $g_1$ fixed. At non-zero $g_2$ the
excited states of adjacent chains are mixed. However, so long as
$g_2$ is small compared to the excitation gap at $g_2=0$, this
mixing will not produce negative energy eigen states.\cite{notee}
Under that condition $|\psi_{2D}>$ continues to be the exact
ground state of $H_{2D}$, and it is separated by an energy gap
from the excited states. Since $|\psi_{1D}>$ does not break
translation symmetry, so does'nt $|\Psi_{2D}>$. 

The remaining task is to collaborate the statements we have made
about $|\psi_{1D}>$. To accomplish this we will make use of a
mapping from the 1D Hamiltonian to a well-studied problem in two
dimensions, namely, the spinless fermions in a strong magnetic
field.

It is well known that when the lowest Landau level (LLL) is $1/3$
filled, and if the interaction between fermions falls off
sufficiently fast as a function of the interparticle distance,
 the ground state is an incompressible quantum fluid.  
It is also well-known that when a particle is confined to the
lowest Landau level the two components of its coordinate no longer
commute. This implies that the dynamics of the fermions
 is effectively that of a 1D system. This intrinsic
one-dimensionality can be made explicit through a mapping of the
LLL problem to one dimension. What has not been appreciated is
that {\it although the time reversal symmetry is explicitly broken
in the 2D problem, the corresponding 1D Hamiltonian is
time-reversal invariant.} Moreover, after the mapping to 1D, the
ground state is an example of a featureless Mott insulator at 1/3
occupation number.

To start, let us consider a spin-polarized two dimensional
electron gas where the electrons interact through a two-body
potential
\be &&H_{int}=V_0\int d^2r d^2r' V(\v r-\v r') \psi^+(\v
r)\psi^+(\v r')\psi(\v r')\psi(\v r)\nn&&V(\v r)=\gr^2\delta(\v
r).\label{pseudo}\ee Although this interaction potential is a
singular function of the coordinates, it has non-singular matrix
element between states in the lowest Landau level. Laughlin's
$\nu=1/3$ wavefucntion \cite{laughlin} is known to be the exact
ground state of this potential \cite{trugman}. The gap to the
lowest excited state is proportional to the parameter $V_0/l_B^4$,
where $l_B$ is the magnetic length.

Next we place the above two dimensional electron gas on a cylinder
with circumference $L$ with the magnetic field perpendicular to
the surface.  In the Landau gauge the LLL basis orbitals are of
the following form \be \phi_n(\v r)={1\over
\sqrt{\pi^{1/2}Ll_B}}e^{i 2n\pi y/L}e^{-(x-2\pi n
l_B^2/L)^2/2l_B^2}.\label{orbit}\ee These orbitals are delocalized
around the cylinder ($y$) while localized along its axis ($x$).
The quantization of the y-momentum in units of $2\pi/L$ explains
why we have a lattice. If we expand the field operator $\psi(\v
r)$, projected to the lowest Landau level, as $ \psi_(\v r)=
\sum_n \phi_n(\v r) c_n$, substitute the result into \Eq{pseudo},
and rescale the coordinates by $L/2\pi$, we obtain the Hamiltonian
\pref{model1} with $\kappa\equiv {2\pi l_B/ L}$, and $g=
{4V_0\over (2\pi)^{3/2}l_B^4}$. Thus $\kappa$ measures the ratio
between the magnetic length and the circumference of the cylinder.
Since $\kappa^{-1}$ controls the hopping lengths of the 1D and 2D
Hamiltonians
, it is clearly desirable to restrict $\kappa>0$. On the other
hand, to preserve the the liquid property of the Laughlin state we
need to keep $L>l_B$ (or $\kappa<2\pi$). To satisfy both
requirements we assume $0<\kappa<1$.



In Landau gauge, the $\nu=1/3$ Laughlin wavefunction on a cylinder
reads \cite{hansson,iso} \be
\psi_{1/3}(\{x_j,y_j\})\propto\prod_{j>k}
\sinh^3\Big({z_j-z_k\over L/\pi}\Big)e^{-\sum_j
x_j^2/2l_B^2},\label{wf2d}\ee where $z=x+iy$. The $\chi(\{n_k\})$
given in \Eq{wfn1} are the coefficients of $\psi_{1/3}$ when
expanded in terms of products of the lowest Landau level orbitals
(\Eq{orbit}) .

At this point we  have mapped $H_{int}$ to $H_{1D}$, and the
Laughlin state $|\Psi_{1/3}>$ to $|\Psi_{1D}>$. Since the above
mapping is a unitary transformation, all known properties of
$|\Psi_{1/3}>$ and $H_{int}$ are preserved. For example, the fact
that $|\Psi_{1/3}>$ describes a quantum liquid at magnetic filling
factor 1/3 translates into the statement that $|\Psi_{1D}>$ is a
quantum liquid at occupation number $1/3$. The fact that
$|\Psi_{1/3}>$  is the non-degenerate ground state of $H_{int}$
implies that $|\Psi_{1D}>$ is the non-degenerate ground state of
$H_{1D}$. Finally, the fact that $H_{int}$ possesses an energy gap
and fractional charge quasiparticles at filling factor $1/3$
implies the same for $H_{1D}$ at lattice occupation 1/3. In this
way we have established that the model presented by $H_{1D}$
possesses a featureless Mott insulating ground state at fractional
occupation number!

By using the 1D to 2D mapping discussed previously it is simple to
map $H_{2D}$ to a 3D Hamiltonian $H'_{int}$ describing coupled
layers: \be &H'_{int}&=\sum_n H_n\nn &H_n&=\int d^2rd^2r'V(\v r-\v
r')\Big[V_1\psi^+_{n}(\v r)\psi^+_n(\v r')\psi_n(\v r')\psi_n(\v
r)\nn&+& V_2\Big(\psi^+_{n+1}(\v r)\psi^+_{n+1}(\v r')\psi_n(\v
r')\psi_n(\v r)+ h.c.\Big)\Big].\ee In the above equation $V(\v
r)$ is the same as that given in \Eq{pseudo}, and $V_{1,2}\propto
g_{1,2}$ respectively. With $V_2=0$ the problem becomes that of
many independent layers. At $\nu=1/3$ each layer is in the liquid
ground state described by $|\Psi_{1/3}>$. The term proportional to
$V_2$ hops a pair of electrons between adjacent layers. As
discussed below \Eq{st2}, so long as $V_2$ is sufficiently smaller
than $V_1$, the ground state is the direct product of the $1/3$
Laughlin liquid in each layer.

It is interesting to note that unlike usual Hamiltonians,
  $H_{1D}$ and  $H_{2D}$ do not have the single-fermion hopping
term.  One can in fact add a single particle hopping  term \be
H'_{1D}= -t e^{-\kappa^2/4}\sum_n~[c^+_{n+1}c_n +
h.c.]\label{hop}\ee to $H_{1D}$ without changing  qualitatively
any of the results. To understand this we note that after mapping
to 2D \Eq{hop} becomes \be -t \int d^2r \cos({2\pi y/L}) \psi^+(\v
r)\psi(\v r),\ee i.e., a periodic potential.  It is clear that so
long as  $|t|$ is much smaller than the energy gap of $H_{int}$,
it will not change the quantum liquid nature of the ground state
and will not collapse the excitation gap. Similar single-particle
terms can by added to
$H_{2D}$. 
Again, so long as the strength of the single particle hopping
terms is sufficiently small, the qualitative results we discussed
above will be preserved.

Models that gives featureless Mott insulating ground state at
fractional occupation number have been difficult to find. What is
special about the models presented in this paper?
To answer this question let us focus on $H_{1D}$. A standard (1D)
interaction potential is local in configuration space (q-space),
i.e.,  \be \hat{V}=\prod_j \int{dq_j} \sum_{i<j}U(q_i-q_j)
|q_1,..,q_N><q_1,..,q_N|~\ee When this type of potential dominates
the Hamiltonian, it favors particles to assume a fixed q-space
configuration. For a repulsive potential such a configuration
often take the form of a regular lattice. $H_{1D}$, on the other
hand, is non-local in q-space and the particles do not necessarily
have to pay a high price in energy to get close. However, there is
a hidden locality in $H_{1D}$ which is revealed not in
configuration space, but in {\it phase space}.

The quantum description of phase space dynamics is most
conveniently done in terms of coherent states. For a single
particle with coordinate $q$ and momentum $p$, the coherent state
$|z>$ satisfies \be\Big[\hat{q}+i
{\lambda^2\over\hbar}\hat{p}\Big]~|z>=~z~|z>,\ee where $z=x+iy$
and $\lambda$ is {\it any} pre-chosen length scale serving to make
$\hat{q}$ and $(\lambda^2/\hbar)\hat{p}$ the same dimension. If
$|q>$ denotes the position eigenstate, it is simple to show that
\be
&&<q|z>~\propto~e^{iqy/\lambda^2}~e^{-(x-q)^2/2\lambda^2}.\label{corr}\ee
and this function defines the transformation between the phase
space and configuration space descriptions. It is interesting to
note if we identify $\lambda$ with the magnetic length and
$q/\lambda^2$ with the y-momentum, the functions given in
\Eq{corr} become precisely the LLL basis orbitals in \Eq{orbit}.
Thus a 2D
quantum mechanical problem defined in the LLL is completely
equivalent to the coherent state representation of a 1D quantum
problem.

Let us go back to the non-local Hamiltonian $H_{1D}$. When
expressed in terms of the coherent state basis, it gets a local
form \be H_{1D}=\sum_{i<j}\int\prod_k{d^2z_k\over 2\pi\lambda^2}
V(z_i-z_j)|z_1,..,z_N><z_1,..,z_N|.\ee 
The $V(z_i-z_j)$ in the above equation is the potential given in
$H_{int}$. This potential prevents particles from getting close
together in the phase space. However due to the non-commutativity
of the phase space coordinates, the particles are not frozen in
any particular configuration. 
We believe the fact that $H_{1D}$ and $H_{2D}$ are linked to
phase-space local potential is the main reason that they have
featureless Mott insulating ground states at fractional occupation
numbers.

In conclusion, we have constructed 1D and 2D lattice models that
exhibit featureless Mott insulating ground state at fractional
occupation number. The Hamiltonians of these models have finite
range interactions, and respect lattice translation and time
reversal symmetries. These Hamiltonians are local in phase space
instead of configuration space. We believe phase space local
interaction could be a missing key for constructing models of this
type of novel insulators.

\begin{acknowledgments}
DHL is supported by DOE grant DE-AC03-76SF00098. JML thanks the
Miller Institute for Basic Research in Science for financial
support and hospitality during his recent visit. Supports from the
Research Council of Norway and the Fulbright foundation are also
acknowledged.
\end{acknowledgments}

\widetext

\begin{references}
\bibitem{band} In a band insulator all low-energy one-particle quantum states
are filled. Such a system is insulating because of the Pauli
exclusion principle. Usually a band insulator has even number of
electrons per unit cell.
\bibitem{notee} We have performed explicit analysis to second order in
$g_2$ to confirm this statement.
\end{references}
\end{document}